\documentstyle[seceq,preprint,epsf,mbf,wrapft]{ptptex}

\newcommand{\D}{\partial}
\newcommand{\be}{\begin{equation}}
\newcommand{\ee}{\end{equation}}
\newcommand{\bea}{\begin{eqnarray}}
\newcommand{\eea}{\end{eqnarray}}
\newcommand{\dt}{\frac{d}{dt}}
\newcommand{\GS}{G_{\!\!\mbox{\tiny S}}}

\newcommand{\CO}{N_{\!\!\mbox{\tiny C}}}
\newcommand{\SR}{\!\!\!/}
\newcommand{\T}{\widehat T}
\newcommand{\M}{\widehat \mu}

\preprintnumber[3.5cm]{
KANAZAWA-99-24\\KUCP-0142\\
October, 1999}

\title{
Application of 
Non-Perturbative Renormalization Group
to Nambu-Jona-Lasinio/Gross-Neveu model 
at Finite Temperature and Chemical Potential
}

\author{%
Hiroaki {\sc Kodama}\footnote{
E-mail address: h-kodama@hep.s.kanazawa-u.ac.jp}
and Jun-Ichi {\sc Sumi}
$^{*,}$
\footnote{%
E-mail address: sumi@phys.h.kyoto-u.ac.jp}
}

\inst{%
Institute for Theoretical Physics, Kanazawa University, Kakuma-machi,
 Kanazawa 920-1192, Japan\\
$^{*}$Department of Fundamental Sciences,
 Faculty of Integrated Human Studies,
 Kyoto University, Kyoto 606-8501, Japan
}

\recdate{
${******~**,~****}$}

\abst{
The chiral phase structure of the Nambu-Jona-Lasinio/Gross-Neveu model 
at finite temperature $T$ and finite chemical potential $\mu$
is investigated
using (Wilsonian) Non-Perturbative Renormalization Group (NPRG). 
In the large $\CO$ limit, the solutions of NPRG with various 
cutoff schemes are shown.
For a sufficiently large ultra-violet cutoff, NPRG results 
coincide with those of Schwinger-Dyson equation
and have little cutoff scheme dependence.
Next, to 
improve the approximation, we incorporate the mesonic fluctuations. 
We introduce the auxiliary fields for mesons, and then 
derive NPRG equation for finite $\CO$. 
The chiral phase structure on $(T,\mu)$ plane beyond the 
leading of $1/\CO$ expansion
is investigated in the sharp cutoff limit. 
$\CO$ dependence of chiral phase diagram is obtained.
}

\begin{document}

\maketitle

\makeatletter
\if 0\@prtstyle
\def\asp{0.5em} \def\bsp{0em}
\else
\def\asp{.3em} \def\bsp{0.3em}
\fi \makeatother

\section{Introduction}
Explorations of the phase structure of hot/dense QCD or its toy models
are of theoretical interest, and it is relevant to the 
heavy ion experiments planned at RHIC and LHC.
At finite temperature and density, the vacuum is expected to move
to the quark gluon plasma (QGP) phase via the phase transition, 
where chiral symmetry is recovered and quarks and
gluons are deconfined.
Since the phase transition to QGP phase is
a non-perturbative phenomenon, 
we need a non-perturbative analysis to understand it. 
Unfortunately, 
there are not so many methods to deal with such problems: 
lattice Monte Carlo 
simulation, Schwinger-Dyson equation (SDE), $1/N$ expansion, $\epsilon$ 
expansion. Non-perturbative renormalization group (NPRG) 
is also one of the methods for such a purpose. 
The effectiveness of the NPRG method in the non-perturbative phenomena 
has been investigated 
by many authors.\cite{rf:RGdsb,order_parameter}

Non-perturbative renormalization group equations describe the response 
from the change of infra-red momentum cutoff $\Lambda$ and can be written down 
exactly. They are the functional differential equations for the Wilsonian 
effective action in which the quantum correction from the high energy 
modes ($p > \Lambda$) are already incorporated. In the practical analysis, 
we approximate the theory space, the functional space of the effective action, 
and project the renormalization group equation (RGE) onto this sub-space. 
By enlarging this sub-space, we can improve the approximations 
systematically. In some cases, the evaluated physical quantities 
converge fast under such a process.\cite{rf:comoving} \ 
This is an
advantageous feature of the NPRG method compared to the asymptotic series, 
$e.g.$ the perturbation theory, $\epsilon$ expansion and $1/N$ expansion. 

In this article we investigate the chiral phase structure of 
Nambu-Jona-Lasinio (NJL)/Gross-Neveu (GN) model
\cite{rf:NJL,rf:GN} \  
at finite temperature and finite chemical potential. 
The exploration of the phase diagram is of fundamental interest.
If the analysis is extended to QCD,  
it is of use for the early universe, the astrophysics of neutron stars 
and the physics of heavy ion collisions. 
We employ NPRG method for the analyses.
It is worth while to examine the applicability of NPRG method,
because there does not exist so many other tools for non-perturbative 
analyses and they do not necessarily work well in any situation.

In lattice Monte Carlo calculation, much exploration has been made 
in the system at finite temperature,\cite{rf:lat_T} \  
while the behavior at finite 
density is much less understood. The non-vanishing chemical potential 
$\mu$ makes fermion determinant to be a complex number, and therefore 
straightforward Monte Carlo methods can not be applied. At present, 
two known candidates avoiding this difficulty, 
$i.e.$ Glasgow algorithm\cite{rf:Glasgow} \ and imaginary chemical potential 
$\mu=i\nu$ method,\cite{rf:Chemical} \ require much larger computer resource 
but unfortunately do not bring any definite results. 
Most of the efforts have been done using SDE, but 
it is difficult to improve the approximation systematically. 

This paper is organized as follows.
In \S\,2 we derive the evolution equation which is one of the 
non-perturbative 
renormalization group equations and explain local potential approximation 
(LPA).\cite{rf:lpa} \  
In \S\,3 we discuss the chiral phase structure in the large 
$\CO$ 
limit, and compare with the results from SDE 
for the fermion mass function $\Sigma(q)$. 
Due to the formal 
equivalence of two methods in the large $\CO$ limit, 
our large $\CO$ result 
should coincide with that of SDE. 
We show consistent results can be obtained in the framework of 
NPRG method.
In \S\,4, we investigate the phase structure 
beyond the large $\CO$ approximation. 
Non-perturbative renormalization group method can approximately 
incorporate the higher order diagrams in $1/\CO$ expansion 
within local potential approximation. 
The $\CO$ dependence of the chiral phase diagram
will be presented
there. 
\S\,5 is devoted to the summary and discussions.

\section{Evolution equation and local potential approximation}

There are three formulations of the non-perturbative renormalization group, 
the Wegner-Houghton equation, the Polichinski equation and the evolution 
equation.\cite{Wilson_Kogut,rf:RGE,rf:FE} \  
They are the continuous version of the block spin 
transformation written in the momentum space and describe the response 
of lowering the infra-red momentum cutoff $\Lambda$. We can find the exact 
form of renormalization group equation (RGE) for the Wilsonian effective 
action and/or the effective average action. The latter one is the one particle 
irreducible part of the Wilsonian effective action. By lowering the cutoff 
we have the effective action at large distance and incorporate the radiative 
corrections from the high energy modes. 
In this article, we employ the evolution equation in Ref.\citen{rf:FE} \ 
and apply to the Nambu-Jona-Lasinio (NJL)/Gross-Neveu (GN) 
model\cite{rf:NJL,rf:GN} \ at finite 
temperature $T$ and 
chemical potential $\mu$. 

The generating functional of connected Green functions is
\be
W_\Lambda[\eta,\bar\eta]=\ln\int D\bar\psi D\psi\;\exp\left\{
-S_{{\rm cut}~\Lambda}^{\rm f}[\bar\psi,\psi]
-S_{\rm bare}[\bar\psi,\psi]+\bar\eta\cdot\psi
-\bar\psi\cdot\eta
\right\},\label{eq:connect}
\ee
where $S_{{\rm cut}~\Lambda}^{\rm f}$ is given by,
\be
S_{{\rm cut}~\Lambda}^{\rm f}[\bar\psi,\psi]
=\int_0^{1/kT}d\tau\int d^{d-1}x~
\bar\psi\Delta^{-1}_f(-i\D,\Lambda)\psi.
\label{eq:cut}
~\footnote{In the following, we choose units such that 
Boltzmann constant $k=1$.}
\ee
Here $\Delta^{-1}_f$ is a cutoff operator and has the property,
\be
\Delta^{-1}_f(p,\Lambda)=C^{-1}(p/\Lambda)
(p\SR-i\mu\gamma_0)\longrightarrow\left\{
\begin{array}{cc}
0 & {\rm for}\quad p\gg\Lambda\\
\infty & {\rm for}\quad p\ll\Lambda.\\
\end{array}
\right.
\ee
At finite temperature, $p_0$ is quantized to the Matsubara frequency,
\bea
\omega_{f,n}=(2n+1)\pi T \quad\mbox{or}\qquad\omega_{b,n}=2n\pi T,
\eea
which is for fermions and bosons, respectively.
Note that, since $S_{{\rm cut}~\Lambda}^{\rm f}$ preserves 
the chiral symmetry, the effective 
action also respects it. 
We choose the cutoff function $C$ as,
\be
C^{-1}(p/\Lambda)=\frac{f^2(p/\Lambda)}{1-f^2(p/\Lambda)},
\quad {\rm with}\quad f(p/\Lambda)=\exp[-a(p/\Lambda)^{2b}],
\label{eq:cutoff_fun}
\ee
Since the parameter $a$ 
can be absorbed in a redefinition of the cutoff $\Lambda$,
we fix it such that $a=0.3$ in this paper. 
The parameter $b$ is the cutoff scheme parameter.
\footnote{In \S\,3, we employ this smooth cutoff regularization
to see cutoff scheme $b$ (in)dependence.}
Taking the derivative of Eq.(\ref{eq:connect}) 
with respect to $\Lambda$
and performing the Legendre transformation 
$\widetilde{\Gamma}_\Lambda[ \bar\psi,\psi]+
\bar\psi\cdot\Delta^{-1}_f\psi=\bar\eta\cdot\psi-\bar\psi\cdot\eta-W_\Lambda
[\bar\eta,\eta]$, we get the evolution equation for the effective average 
action, 
\be
\Lambda\frac{d}{d\Lambda}\widetilde\Gamma_\Lambda[ \bar\psi,\psi]
=-\frac{1}{2}{\bf str}\left[\Lambda\frac{d}{d\Lambda}\Delta^{-1}_f\left(
\Delta^{-1}_f+
\widetilde\Gamma_\Lambda^{(2)}\right)^{-1}\right],
\ee
where ${\bf str}$ is super-trace which involves momentum (or coordinate)
integration, Matsubara summation, spinor summation and color summation. 
$\widetilde\Gamma_\Lambda^{(2)}$ is a second (functional) derivative 
with respect to 
the fields $\Phi=(\psi^{\rm T},\bar\psi)$ $i.e.$
\be
\left(\widetilde\Gamma_\Lambda^{(2)}\right)_{xy}\equiv
\frac{\overrightarrow\delta}{\delta\Phi_{x}}
\widetilde\Gamma_\Lambda[\Phi]
\frac{\overleftarrow\delta}{\delta\Phi^{\rm T}_{y}}\;.
\ee
This RGE possesses the exact information about the response of the 
effective average action 
to the coarse graining. However, it is a functional differential equation 
and we can not solve it without approximation.
As a first step of our approximation, 
we neglect higher derivative terms and keep $Z$ factor to be unity. 
It is called the local potential approximation (LPA).
\cite{rf:lpa} \ 
This is the leading order of the derivative expansion.
\cite{rf:de} \  
\begin{figure}
    \epsfxsize=10cm
    \centerline{\epsfbox{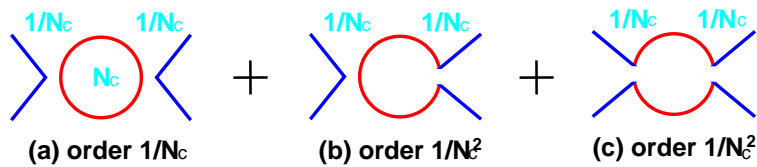}}
\caption{The Feynman diagrams of four-fermi $\beta$ functions 
incorporated in the LPA. 
The leading contribution in $1/N_{\mbox{c}}$ 
expansion corresponds to the 
first diagram (a). Lines denote the flow of color indices.}
\label{fig:1}
\end{figure}
Since the NPRG preserves the homogeneous global symmetries, 
the effective average action $\widetilde\Gamma_\Lambda$ also respect it. 
So the operator space is restricted to that of the chiral invariants.
For example, the independent chiral invariant four-fermi operators are
given as; 
${\cal O}_1=(\bar{\psi}\psi)^2+(\bar{\psi}i\gamma_5\psi)^2$,  
${\cal O}_2=(\bar{\psi}\gamma_i\psi)^2+(\bar{\psi}\gamma_i\gamma_5\psi)^2$
and 
${\cal O}_3=(\bar{\psi}\gamma_0\psi)^2+(\bar{\psi}\gamma_0\gamma_5\psi)^2$
\footnote{We consider only the case with a single flavor.}.
Here color indices are omitted 
and $i$ is the spatial index running over $i=1,2,3$. 
In the LPA, the effective average action is
\be
\widetilde\Gamma_\Lambda[\bar\psi,\psi]=\int_0^{1/T}d\tau\int d^{d-1}x\left\{
\bar\psi(i\partial\SR-i\mu\gamma_0)\psi
-\sum_i\frac{G_i}{2\CO}{\cal O}_i
+\cdots
\right\},
\ee
where $\CO$ is the number of colors. 
In the LPA, the four-fermi operators do not 
receive any corrections from the multi-fermi operators other than 
the four-fermi 
operators. 
The Feynman diagrams corresponding to the $\beta$ functions of 
the four-fermi coupling constants are drown in Fig.~\ref{fig:1}.
In the large $\CO$ limit, 
the Feynman diagrams (b) and (c) in Fig.~\ref{fig:1} 
don't contribute to our $\beta$ functions.

\section{Phase structure in the large ${N_{\mbox{c}}}$ limit}

\begin{figure}[hbt]
\parbox[t]{63mm}{
\epsfxsize=0.471\textwidth
\epsfbox{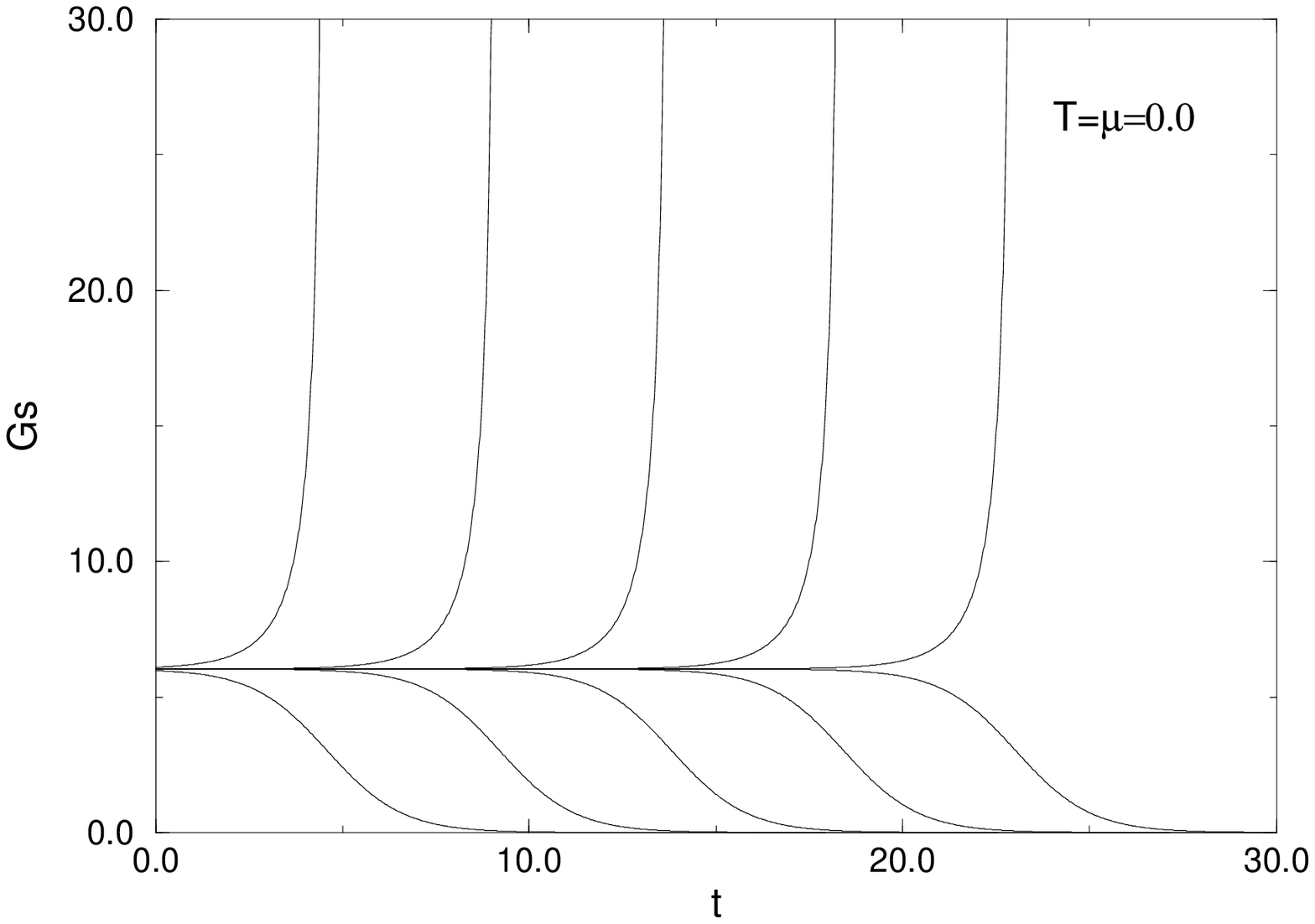}
\caption{RG flow diagram for four-fermi coupling 
${\widehat G}_{\mbox{s}}$
of three dimensional GN model in the large $N_{\mbox{c}}$ limit
at zero temperature and zero chemical potential.
The cutoff scheme is $b=2.4$.}
\label{fig:2}
}
\hspace*{8mm}
\parbox[t]{63mm}{
\epsfxsize=0.471\textwidth
\epsfbox{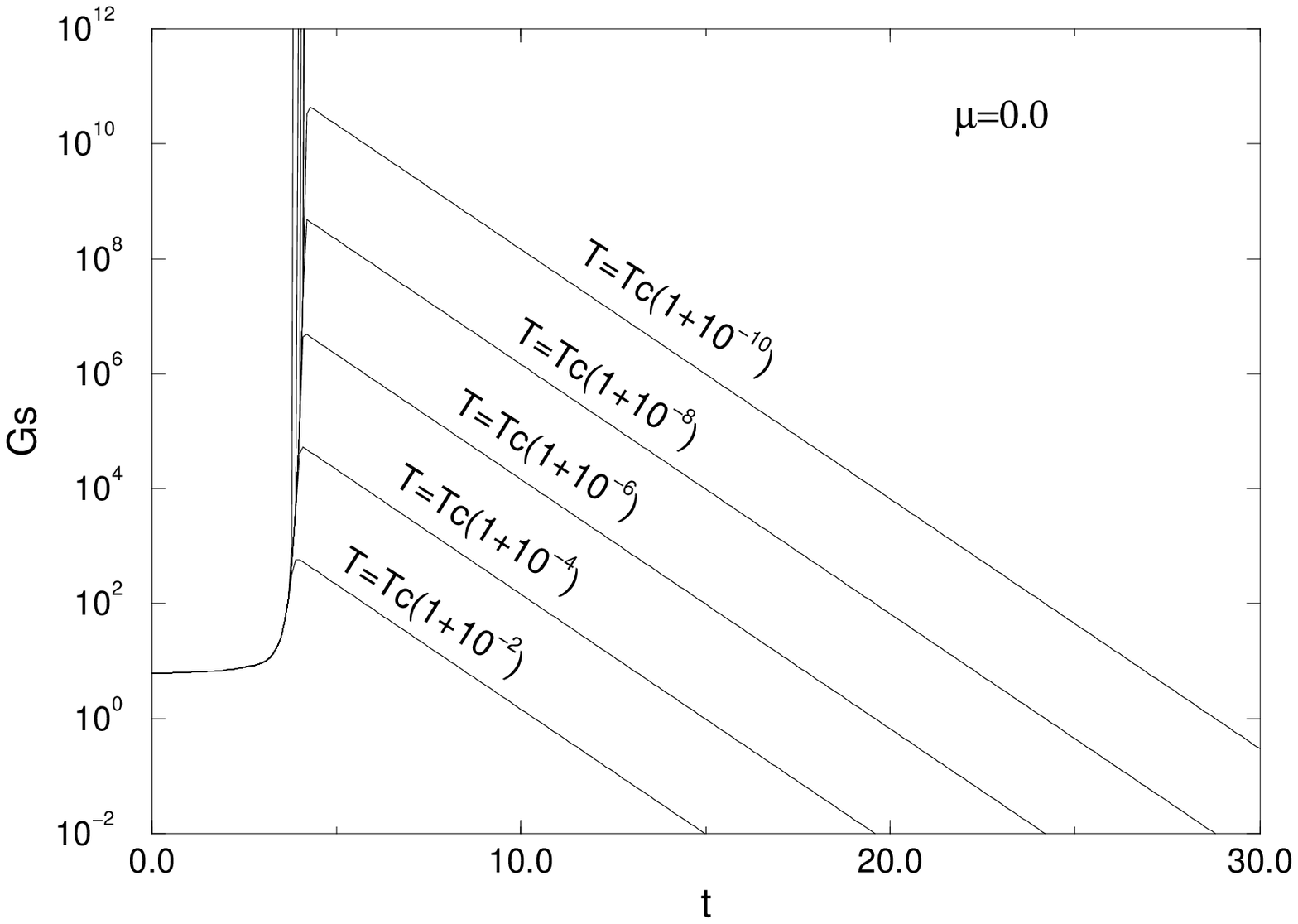}
\caption{RG flow diagram for four-fermi coupling ${\widehat G}_{\mbox{s}}$
of three dimensional GN model in the large $N_{\mbox{c}}$ limit
at finite temperature and zero chemical potential. 
The cutoff scheme is $b=2.4$. }
\label{fig:3}
}
\end{figure}

In this section, we explore the phase structure of the Gross-Neveu
model\cite{rf:GN} \ at finite 
temperature and chemical potential in the large $\CO$ limit. 
We attempt to apply the NPRG method to GN model at $T\ne0$ and $\mu\ne0$, 
and show NPRG reproduces consistent results with
SDE.\cite{Inagaki} \  
In the large $\CO$ limit, 
only the first diagram (a) in Fig.~\ref{fig:1} 
contributes to the $\beta$ function of the four-fermi couplings. 
The $\beta$ function of the scalar four-fermi coupling 
$\GS\equiv G_1$
is a function of $\GS$ alone,
\be
\dt{\widehat \GS }=-(d-2){\widehat \GS }+2{\widehat \GS }^2
I(a,b~;{\widehat T},{\widehat \mu}),\label{eq:rgs3}
\ee
where $t$ is the cutoff scale parameter $i.e.$ $\Lambda=\Lambda_0\exp(-t)$. 
A profile of the threshold function $I(a,b;{\widehat T},{\widehat\mu})$ 
is given in Appendix A. 
The characters with hat are the dimensionless coupling constant, $e.g.$ 
${\widehat \GS }=\GS /\Lambda^{d-2}$. 
The first term of the right hand side (RHS) 
of Eq.(\ref{eq:rgs3}) corresponds to 
the canonical scaling and 
the second one to the radiative correction. 
At zero temperature and zero chemical potential, we find the two-phase 
structure by solving the RGE numerically. 
The RG flow diagram is shown in Fig.~\ref{fig:2}.
There are two phases divided by critical coupling, 
the strong coupling phase and the weak coupling phase. 
In the strong coupling phase, the four-fermi coupling constant blows up to 
the infinity at finite scale $t$. 
By evaluating 
the effective potential of meson fields, 
we can recognize that the chiral symmetry 
is spontaneously broken in this phase.\cite{order_parameter}
In the weak coupling phase, 
the four-fermi coupling goes to zero and
chiral symmetry is not broken. 

Let us discuss hot and dense matter. 
The broken chiral symmetry at zero temperature and 
zero chemical potential is 
restored at some critical temperature $T_c$ and/or 
critical chemical potential $\mu_c$. 
We show the RG flow of the four-fermi coupling 
constant at finite temperature in Fig.~\ref{fig:3}.
There is the critical temperature, below which
the four-fermi coupling constant blows up to the infinity.
On the other hand, above the critical temperature
the four-fermi coupling turns to decrease exponentially. 

\begin{figure}[hbt]
\parbox[t]{63mm}{
\epsfxsize=0.471\textwidth
\epsfbox{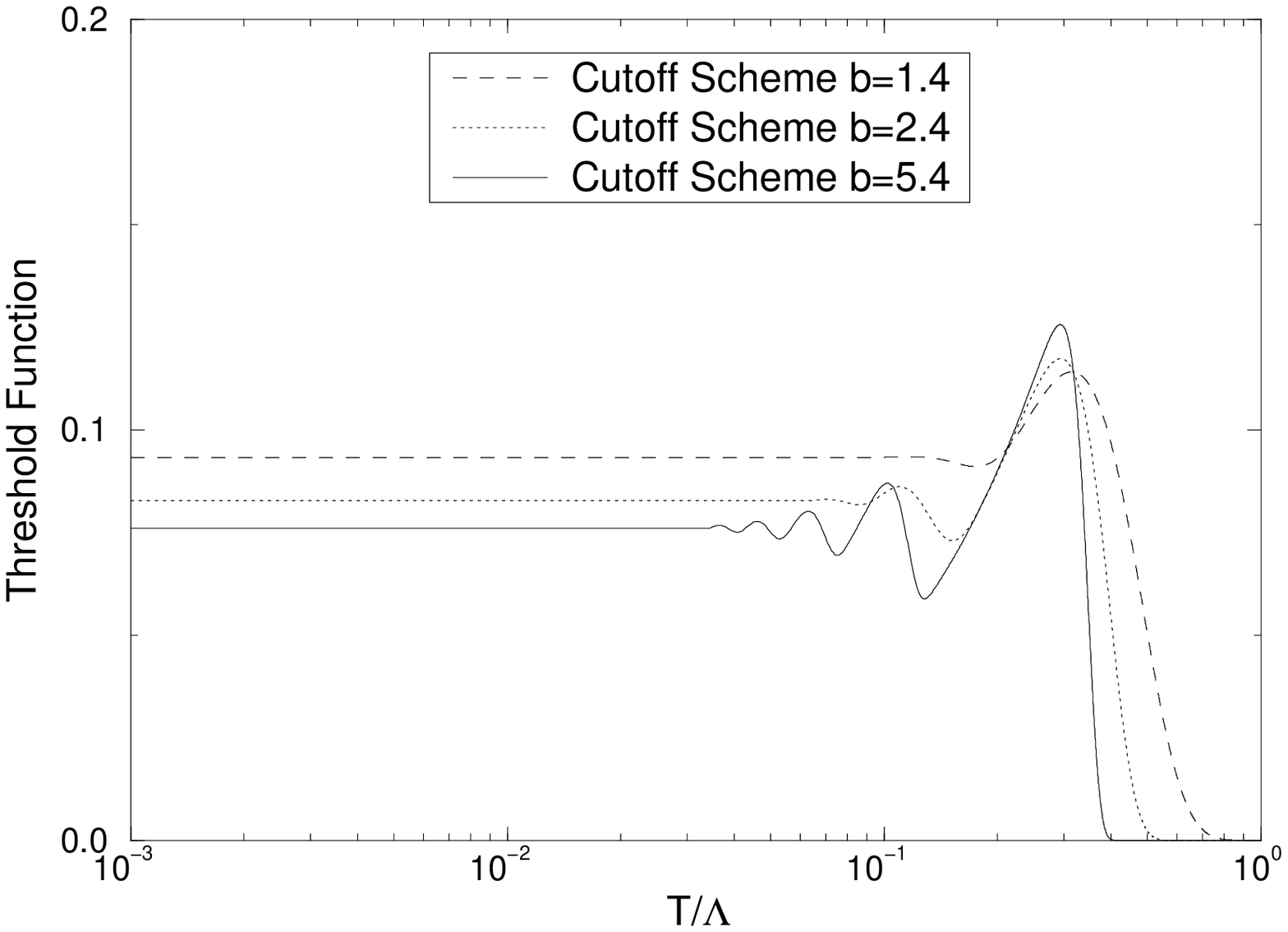}
\caption{Temperature dependence of the threshold function $I$ at $\mu=0$ 
with various cutoff schemes in three dimensional GN model.}
\label{thresh}
}
\hspace*{8mm}
\parbox[t]{63mm}{
\epsfxsize=0.471\textwidth
\epsffile{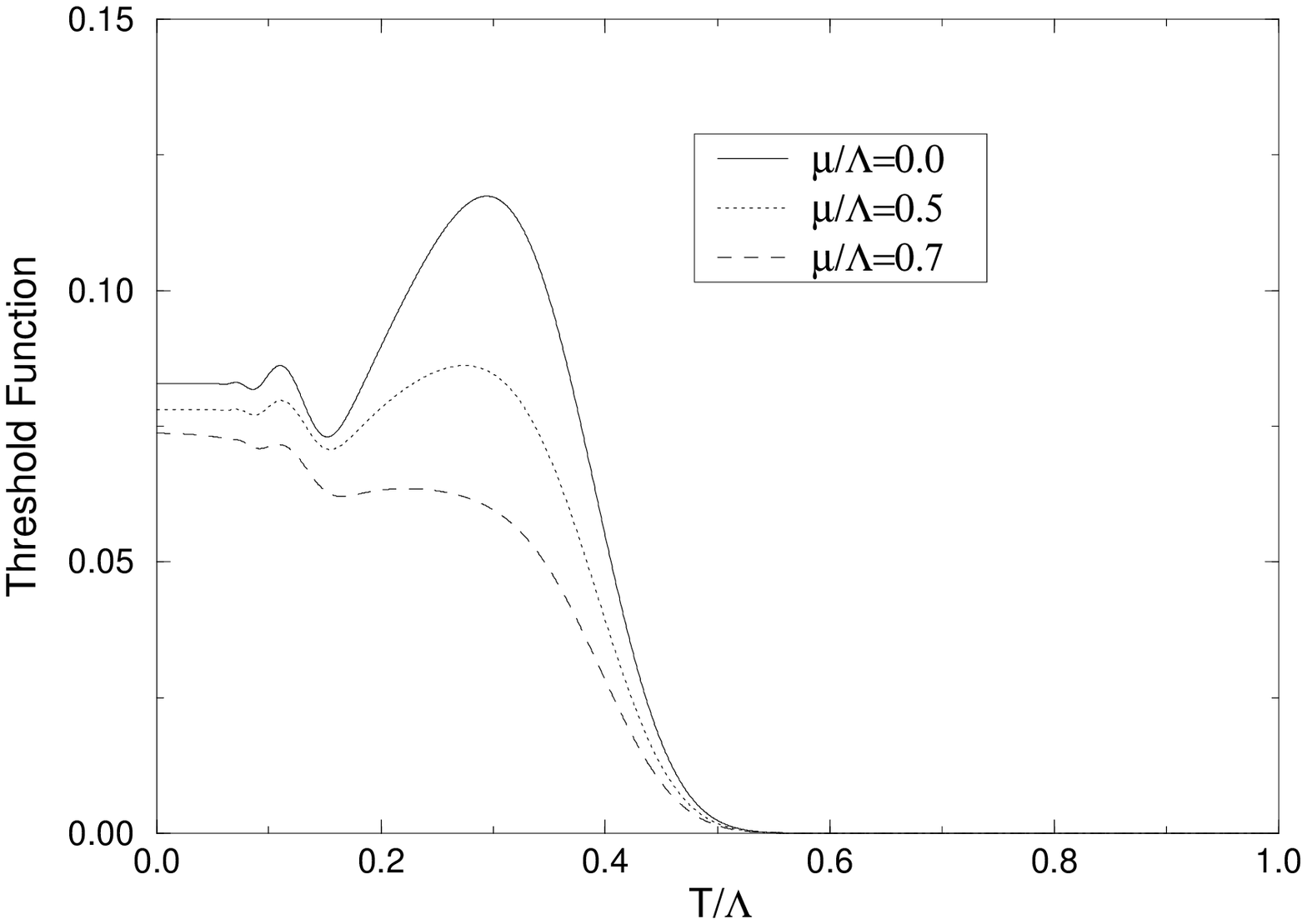}
\caption{Temperature dependence of the threshold function $I$ at $\mu\ne0$ 
in three dimensional GN model. The cutoff scheme is $b=2.4$.}
\label{threshould_D}
}
\end{figure}

The temperature/chemical potential dependence of 
various quantities can be found 
by solving RGE with the same initial condition as 
that of $T=\mu=0$. 
There are the critical 
temperature/chemical potential, above which the four-fermi coupling tends to 
zero, so that chiral symmetry is restored. 
We can estimate the critical temperature and the critical chemical potential 
by solving the RG flow equation 
for the scalar four-fermi coupling constant with 
some fixed initial condition, or equivalently bare coupling constant. 
The initial condition should be given at $\Lambda_0\to\infty$.
However, then we must calculate contributions of infinite number of 
Matsubara modes since $T/\Lambda={\widehat T}$ goes to zero.
\footnote{As seen from the explicit expression of threshold function in 
Appendix A, the contributions of high Matsubara modes are 
suppressed exponentially for a finite ${\widehat T}$.
So at some high Matsubara modes, they become to make no contribution
within the accuracy of numerical computation.
In ${\widehat T}\to 0$ limit, however, infinite number of 
Matsubara modes contribute.}
In the practical analysis, we solve the RG flow equations with 
the common initial condition at the sufficiently large but finite 
ultra-violet cutoff $\Lambda_0$. 
If $\Lambda_0$ is sufficiently large compared with $T$ and $\mu$,
the solutions will reach the scaling region where 
the renormalized information $i.e.$ $\Lambda_0\to\infty$ limit
is obtained.
This corresponds to
tuning the bare four-fermi coupling constant to the critical one : 
$\widehat\GS|_{\Lambda=\Lambda_0}=\widehat\GS^\star+\delta\widehat\GS$, 
$\delta\widehat\GS\to 0$, where $\widehat\GS^\star$ is the critical 
coupling constant at $T=\mu=0$.
For sufficiently large $\Lambda_0$
(or sufficiently small $\delta\widehat\GS$), 
the critical temperature and critical chemical potential 
proportionally depend on the common factor
$\delta{\widehat\GS}^{-1/(2-d)}$, $i.e.$ 
$T_{\mbox{c}}(\delta \widehat\GS)\propto\delta{\widehat\GS}^{-1/(2-d)}$, 
$\mu_{\mbox{c}}(\delta \widehat\GS)\propto\delta{\widehat\GS}^{-1/(2-d)}$.
\footnote{%
Other physical quantities, such as the fermion effective mass $m_0$ 
at $T=\mu=0$ also proportionally depend on 
the factor $\delta{\widehat\GS}^{-1/(2-d)}$.
For the detailed explanation, see Appendix B.} 
Thus the ratio of these quantities is independent of 
the initial value of the four-fermi coupling
for the scaling region. 
The practical problem in the calculations is 
how large $\Lambda_0$ is needed 
to reach the scaling region.
In Figs.~\ref{thresh} and \ref{threshould_D}, 
temperature and chemical potential dependence of threshold function $I$ 
is shown.
At small dimensionless temperature $T/\Lambda < 0.03$, threshold functions 
seem almost constant 
\footnote{
They are the values at $T=\mu =0$.}
and have almost no temperature dependence.
So we expect to reach the scaling region if we set the 
initial condition at $\Lambda_0 > 33~T$.
However, in practice, we must check whether the physical quantities
scale correctly or not, since 
it may depend on the quantities to calculate.
We perform the scaling check by calculating a ratio of
critical temperature (or chemical potential) and $m_0$.
Here $m_0$ is dynamical fermion mass at $T=\mu=0$.

\begin{wrapfigure}{r}{\halftext}
\epsfxsize=\halftext
\epsfbox{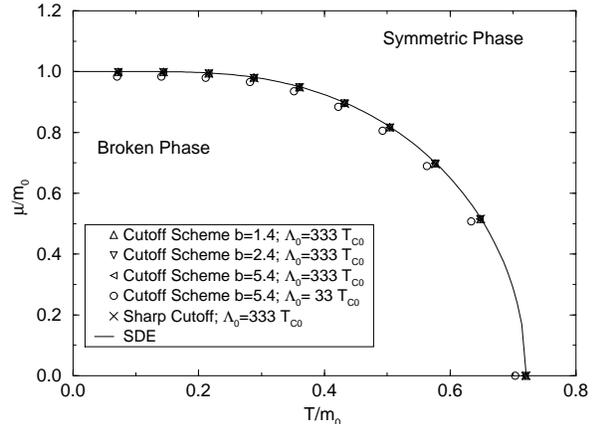}
\caption{Chiral phase structure 
of three dimensional GN model
at finite temperature and chemical potential
in the large ${N_{\mbox{c}}}$ limit.
Temperature and chemical potential are 
normalized by dynamical fermion mass $m_0$
at $T=\mu=0$.
The dependence of both
the cutoff scheme and ultra-violet cutoff $\Lambda$
in the case of NPRG method
Eq.~(3.2) 
and also the results of SDE are shown.
The phase transition is of second order all along the critical line.
}
\label{TDplaneD3I}
\end{wrapfigure}

We show the results in Fig.~\ref{TDplaneD3I}.
If we employ the ultra-violet cutoff $\Lambda_0=333~T_{\mbox{c\tiny{0}}}$,
the critical points 
$(T_{\mbox{c}}/m_0,\mu_{\mbox{c}}/m_0)$ almost coincide with those
of SDE.
Here $T_{\mbox{c\tiny{0}}}$ is a critical temperature
at $\mu=0$.
There is little cutoff scheme dependence 
using this ultra-violet cutoff $\Lambda_0$.
With the ultra-violet cutoff $\Lambda_0=333~T_{\mbox{c\tiny{0}}}$
these quantities reach the scaling region.
On the other hand, 
if we choose the ultra-violet cutoff $\Lambda_0=33~T_{\mbox{c\tiny{0}}}$,
the critical points 
$(T_{\mbox{c}}/m_0,\mu_{\mbox{c}}/m_0)$ deviates from those
of SDE a little.
$\Lambda_0=33~T_{\mbox{c\tiny{0}}}$ is not sufficient 
to obtain a scaled quantities.
We can calculate scaled critical temperature/chemical potential
by setting the initial condition at $\Lambda_0\ge 333~T_{\mbox{c\tiny{0}}}$.
We also plot the results in the sharp cutoff case 
$i.e.$ $b=\infty$ in Fig.~\ref{TDplaneD3I}.
\footnote{The detailed explanation in the sharp cutoff case
is given in \S\,4.}

It should be noted that 
above analyses are based on the flow equation for 
$1/\widehat\GS$ rather than Eq.(\ref{eq:rgs3}) for a technical reason. 
In the large $\CO$ limit, 
$1/\widehat\GS$ coincides with the mass squared of the meson fields. 
Multiplying $-1/{\widehat \GS}^2$ to RGE~(\ref{eq:rgs3}), it
can be rewritten as
\be
\dt\left(\frac{1}{\widehat \GS }\right)=(d-2)\cdot\frac{1}{\widehat \GS }-2
I(a,b~;{\widehat T},{\widehat \mu}).\label{eq:rgs3_2}
\ee
Two phases, chiral broken phase and symmetric phase on $(T,\mu)$ plane 
can be distinguished by the behavior of the solution of eq.(\ref{eq:rgs3_2}). 
If $1/{\widehat \GS}$ goes to the negative value at sufficiently large $t$,
it is the strong coupling (broken) phase. 
If $1/{\widehat \GS}$ tends to the positive infinity,
it is the weak coupling (symmetric) phase.
Although Eq.(\ref{eq:rgs3_2}) and Eq.(\ref{eq:rgs3}) are 
essentially identical,
there emerge difference on the analyses.
If chemical potential ${\widehat \mu}$ is large enough, 
the threshold function $I$ can take negative value.
So even if $1/{\widehat \GS}$ become negative, $1/{\widehat \GS}$ 
may turn back to the positive value due to the negative $I$. 
Thus it should be regarded as weak coupling phase.
With RGE Eq.(\ref{eq:rgs3}), 
this behavior can not be detected 
because once ${\widehat \GS}$ blow up to the infinity, 
then by mistake we may identify it as the strong coupling phase. 
Indeed, for a large $\mu$, it gives an inaccurate (about 10\%) result.
The analysis using $1/{\widehat \GS}$ criticality is a more favorable method
and we may say, it is a sort of ``environmentally friendly
renormalization group''.\cite{environment} \ 

In Ref.~\citen{Inagaki}, the chiral phase structure of GN model 
in $2\le d<4$ dimensions were calculated using 
SDE.
The effective potentials are calculated in the leading order
of the $1/\CO$ expansion. 
It is known that in $2\le d<3$ chiral phase transition is of first order 
at low temperature and large chemical potential, 
and of second order in other region of critical line. 
In $3\le d<4$, 
the phase transition is of second order all along the critical line.

\begin{figure}[hbt]
\parbox[t]{63mm}{
\epsfxsize=0.471\textwidth
\epsfbox{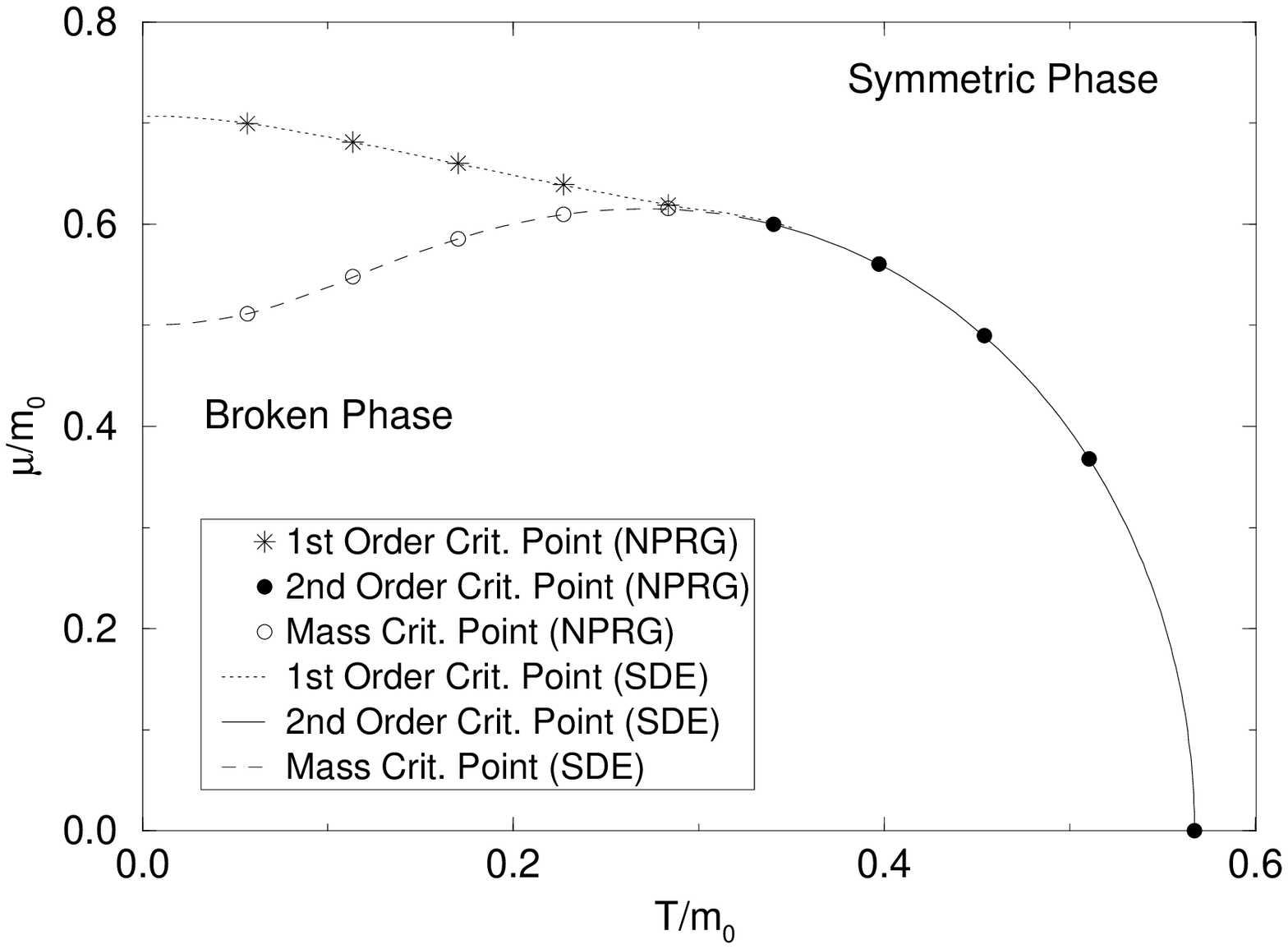}
\caption{Chiral phase structure 
of two dimensional GN model
at finite temperature and chemical potential
in the large ${N_{\mbox{c}}}$ limit.
Temperature and chemical potential are 
normalized by dynamical fermion mass $m_0$
at $T=\mu=0$.
The symbols are the results using NPRG method.
Here cutoff scheme is $b=1.4$.
Ultra-violet cutoff $\Lambda_0$ is $333~T_{\mbox{c\tiny{0}}}$.
The lines are the results using SDE.
Phase boundary consists of the first order critical line
and the second order one.
The dashed line and the open circles have nothing to do
with the phase transition.
}
\label{TDplaneD2}
}
\hspace*{8mm}
\parbox[t]{63mm}{
\epsfxsize=0.471\textwidth
\epsfbox{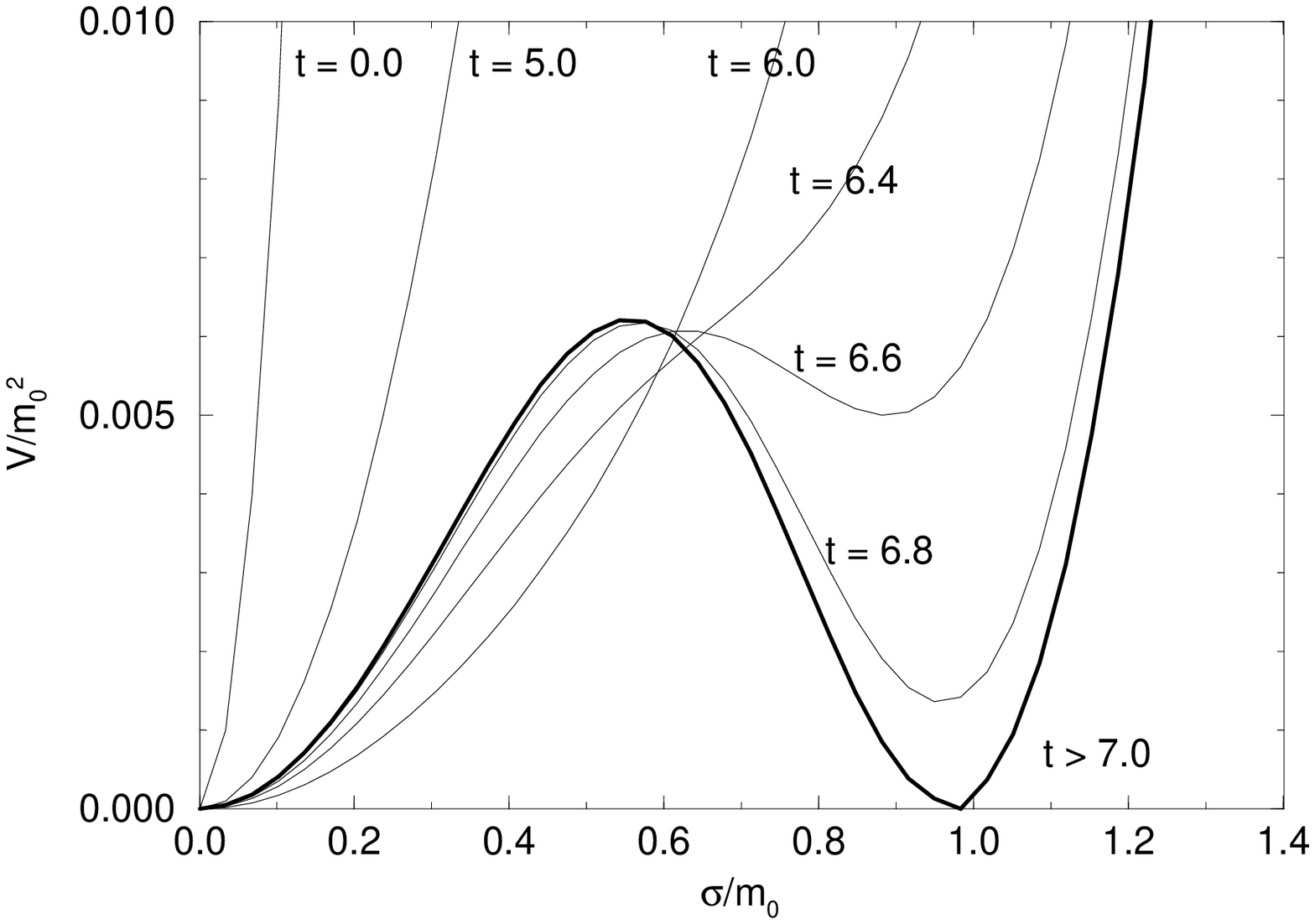}
\caption{RG history of the effective potential of two dimensional
GN model on the first order phase boundary 
in the large ${N_{\mbox{c}}}$ limit.
We choose $T/m_0=0.113,\mu/m_0=0.682$.
The effective potential and the field expectation value are 
normalized by 
the dynamical fermion mass $m_0$ at $T=\mu=0$.
Cutoff scheme here is $b=1.4$.
Ultra-violet cutoff $\Lambda_0$ is $333~T_{\mbox{c\tiny{0}}}$.
}
\label{Pot_1st}
}
\end{figure}

Using NPRG we can also derive 
consistent phase boundaries in any space-time dimensions. 
For example, 
the chiral phase structure of two dimensional GN model 
\footnote{
In two dimensions continuous symmetry can not be spontaneously 
broken.\cite{no-Goldstone} \ 
So we consider the phase transition with a discrete chiral symmetry
$\psi\to\gamma_5\psi$ in two dimensions.
}
is shown in Fig.~\ref{TDplaneD2}.
In two dimensions, there is the first order critical line as well as
the second order one.
The phase boundary can be found only by exploring the effective
potential $V$ of the collective coordinate $\sigma$, which is introduced 
in the next section.
The second order critical point is 
where $\partial^2 V/\partial\sigma^2|_{\sigma=0}=0$ is satisfied,
while the first order critical point is 
where the value of the effective potential at the second minimum
is equal to that at the first minimum at the origin.
The stars in Fig.~\ref{TDplaneD2} stands for the first order 
chiral phase boundary and it terminates at the tricritical point.
The filled circles and the open circles are 
where the signs of meson masses at the origin 
$<\sigma>=0$ are changed and they correspond to
the $1/{\widehat\GS}$ critical line.
The filled circles are identical to 
the second order chiral phase boundary, but the open circles
have nothing to do with the phase transition.
The effective potential 
of two dimensional GN model
has a gap structure on the first order phase boundary.
An example of the RG evolution of the effective potential at a 
first order critical point is shown in Fig.~\ref{Pot_1st}.

\section{Phase structure beyond the $1/{N_{\mbox{c}}}$ leading}

In the previous section, we explored chiral phase structure in $(T,\mu)$ plane 
in the large $\CO$ limit. 
In this section, we improve the approximation 
beyond the $1/\CO$ leading in four dimensional NJL model. 
Since we are interested 
not only in the phase boundary but also in the order of phase transition,
the CJT effective potential\cite{rf:CJT} should be investigated. 
Let us start from the definition of the partition function,
\be
Z_\Lambda[J]=\int D{\bar \psi}D\psi~\exp \left\{-S_{{\rm cut}~\Lambda}^{\rm f}
-S_{\rm bare}+\bar\eta\cdot\psi
-\bar\psi\cdot\eta+\Sigma_\sigma\cdot\bar\psi\psi
+\Sigma_\pi\cdot\bar\psi i\gamma_5\psi
\right\},\label{eq:pathZ}
\ee
where $S_{{\rm cut}~\Lambda}^{\rm f}$, $S_{\rm bare}$ 
and $J$ are the cutoff 
action defined in Eq.(\ref{eq:cut}), the bare action and 
the sources $J=\{\eta,{\bar \eta},\Sigma_\sigma,\Sigma_\pi\}$, respectively. 
Here the bare action is taken to be,
\be
S_{\rm bare}=\int d^4x\left\{
\bar\psi(i\D\SR-i\mu\gamma_0)\psi-\frac{\GS}{2\CO}
\left[(\bar\psi\psi)^2+(\bar\psi i\gamma_5\psi)^2
\right]
\right\}.\label{eq:bareaction}
\ee 
We introduce the auxiliary fields $\sigma,\pi$ corresponding to the composite 
operators ${\bar \psi}\psi,~{\bar \psi}i\gamma_5\psi$
by the following Gaussian integral, 
\be
{\widehat Z}_\Lambda=\int D\sigma D\pi~e^{
-S_{{\rm cut}~\Lambda}^{\rm b}-\int d^4x~\left\{
\frac12P^{-1}\left(\sigma-\Sigma_\sigma-P{\bar \psi}\psi\right)^2
+\frac12P^{-1}\left(\pi-\Sigma_\pi
-P{\bar\psi}i\gamma_5\psi\right)^2\right\}},\label{eq:auxiliary}
\ee
where $S_{{\rm cut}~\Lambda}^{\rm b}$ is the cutoff action 
of meson fields and given by,
\be
S_{{\rm cut}~\Lambda}^{\rm b}=\int d^4x~
\frac12 \left(\D_\mu\sigma~ C^{-1}(-i\D/\Lambda)~\D_\mu\sigma
+\D_\mu\pi~ C^{-1}(-i\D/\Lambda)~\D_\mu\pi\right).
\ee
Here $C^{-1}$ is the cutoff function defined in
Eq.(\ref{eq:cutoff_fun}).
Note that, since ${\widehat Z}_\Lambda$ depends on the fields $\bar\psi,~\psi$ 
as well as on the sources $\Sigma_\sigma,~\Sigma_\pi$, the partition 
function is deformed by the insertion of ${\widehat Z}_\Lambda$ except at 
$\Lambda=0$. Inserting ${\widehat Z}_\Lambda$ into the path-integral 
of Eq.(\ref{eq:pathZ}),
the partition function becomes
\bea
&&Z'_\Lambda[J]=\int D{\bar \psi}D\psi D\sigma D\pi
~e^{-S_{{\rm cut}~\Lambda}^{\rm f}-S_{{\rm cut}~\Lambda}^{\rm b}
-S_{\rm bare}+\bar\eta\cdot\psi
-\bar\psi\cdot\eta+P^{-1}\Sigma_\sigma\cdot\sigma
+P^{-1}\Sigma_\pi\cdot\pi
}\nonumber\\
&&\qquad\qquad\qquad
\times e^{-\int d^4x~\left\{
\frac12P^{-1}(\sigma^2+\pi^2)-\bar\psi(\sigma+\pi i\gamma_5)\psi
+\frac12P\left[({\bar \psi}\psi)^2+({\bar \psi}
i\gamma_5\psi)^2\right]+\frac{1}{2P}({\Sigma_\sigma}^2+{\Sigma_\pi}^2)
\right\}}\nonumber\\
&&\qquad\quad=\int D{\bar \psi}D\psi
~e^{-S_{{\rm cut}~\Lambda}^{\rm f}
-S_{\rm bare}'+\bar\eta\cdot\psi
-\bar\psi\cdot\eta+\Sigma_\sigma\cdot\bar\psi\psi
+\Sigma_\pi\cdot\bar\psi i\gamma_5\psi
},
\eea
where $S_{\rm bare}'$ is given by,
\bea
&&S_{\rm bare}'=S_{\rm bare}-
\frac12\int d^4x\Bigl\{
\left({\bar \psi}\psi+P^{-1}\Sigma_\sigma\right)
\left[(P^{-1}+C^{-1}\Box 
)^{-1}-P\right]
\left({\bar \psi}\psi+P^{-1}\Sigma_\sigma\right)\nonumber\\
&&\qquad\quad
+\left({\bar \psi}i\gamma_5\psi+P^{-1}
\Sigma_\pi\right)
\left[(P^{-1}+C^{-1}\Box 
)^{-1}-P
\right]
\left({\bar \psi}i\gamma_5\psi+P^{-1}
\Sigma_\pi\right)\Bigr\}.
\eea
In the limit $\Lambda\to 0$, $Z'_\Lambda=Z_\Lambda$ 
except for the field independent constant.
We modify the generating functional $W_\Lambda[J]=\ln Z'_\Lambda[J]$
as $W'_\Lambda=W_\Lambda+1/(2P)\cdot 
(\Sigma_\sigma^2+\Sigma_\pi^2)$.
Physically meaningful quantities
are not affected by this modification.\cite{rf:Muta} 
Indeed, no vacuum expectation values is changed by adding 
an arbitrary polynomial of the sources to the generating functional,
$e.g.$ in our case,
\be
<\bar\psi\psi>=
\left. \frac{dW_\Lambda}{d\Sigma_\sigma}\right|_{\Sigma_{\sigma,\pi}=0}
=\left. \frac{dW'_\Lambda}{d\Sigma_\sigma}\right|_{\Sigma_{\sigma,\pi}=0}
-\left. P^{-1}
\Sigma_\sigma\right|_{\Sigma_{\sigma,\pi}=0}
=\left. \frac{dW'_\Lambda}{d\Sigma_\sigma}\right|_{\Sigma_{\sigma,\pi}=0}.
\ee
We rescale the sources as
$J'=\{\eta,{\bar \eta},\Sigma'_\sigma,\Sigma'_\pi\}
=\{\eta,{\bar \eta},P^{-1}\Sigma_\sigma,P^{-1}\Sigma_\pi\}$. 
The initial condition of RGE at the ultra-violet cutoff $\Lambda=\Lambda_0$
can be given after the Legendre transformation,
$\Gamma^{\mbox{\tiny CJT}}_{\Lambda}[ \bar\psi,\psi,\sigma,\pi]=
\bar\eta\cdot\psi-\bar\psi\cdot\eta
+\Sigma'_\sigma\cdot\sigma+\Sigma'_\pi\cdot\pi
-W'_{\Lambda}[J']$. We have
\bea
\Gamma^{\mbox{\tiny CJT}}_{\Lambda_0}&=&
S_{\rm bare}+S_{{\rm cut}~\Lambda_0}^{\rm f}+S_{{\rm cut}~\Lambda_0}^{\rm b}
+\frac12 \int d^4x P^{-1}\left\{
\left(\sigma-P
{\bar \psi}\psi\right)^2+
\left(\pi-P{\bar \psi}i\gamma_5\psi\right)^2\right\}\nonumber\\
&=&
S_{{\rm cut}~\Lambda_0}^{\rm f+b}
+\int d^4x\left\{{\bar\psi}(i\partial\SR-i\mu\gamma_0
-\sigma-\pi i\gamma_5)\psi
+\frac{\CO}{2\GS}(\sigma^2+\pi^2)\right\},
\label{eq:bare}
\eea
where $S_{{\rm cut}~\Lambda}^{\rm f+b}
=S_{{\rm cut}~\Lambda}^{\rm f}+S_{{\rm cut}~\Lambda}^{\rm b}$. 
On the last line of Eq.(\ref{eq:bare}), we took $P=\GS/\CO$.
By subtracting the cutoff action we define the effective 
averaging action as :
$\widetilde \Gamma_\Lambda
[\psi,{\bar \psi},\sigma,\pi](=\int d^4x{\cal L}_\Lambda)
\equiv\Gamma^{\mbox{\tiny CJT}}_{\Lambda}
-S_{{\rm cut}~\Lambda}^{\rm f+b}$. 
In this paper, 
we approximate it as follows : 
\bea
&&{\cal L}_\Lambda=
{\bar \psi}(Z_{f\tiny{0}}~i\D_0-i\mu)\gamma_0\psi
+Z_{f}{\bar \psi}i\D_i \gamma_i\psi
+\frac12 \CO Z_{b\tiny{0}}((\D_0\sigma)^2+(\D_0\pi)^2) 
\nonumber\\
&&\qquad
+\frac12 \CO Z_{b}((\D_i\sigma)^2+(\D_i\pi)^2)
+y{\bar \psi}(\sigma+i \gamma_5 \pi)\psi
+\CO V(\sigma^2+\pi^2).
\eea
There are a yukawa interaction and an effective potential $V$
which is a function of $\sigma^2+\pi^2$.
Here, by a lack of Lorentz symmetry, the wave-function renormalization 
factors of temporal derivative and those of spatial one are different 
each other. 

The RG flow equation for the effective average action 
$\widetilde \Gamma_\Lambda$ can be derived as done in \S\,2. 
In our approximation, the RGE becomes 
a partial differential equation for the effective potential 
$V(\rho)$ where $\rho=1/2\cdot(\sigma^2+\pi^2)$. 
If we employ the smooth cutoff function $C(q/\Lambda)$ 
as done in the previous section,
we have to perform the momentum integration with respect to 
the spatial momenta
numerically for each Matsubara modes. 
It is very time-consuming and not practical.
As investigated in the previous section, the results 
do not depend on the cutoff scheme for a sufficiently large
ultra-violet cutoff $\Lambda_0$ at least in the large $\CO$ limit.
Therefore it is more promising way to take the sharp cutoff limit 
of the RG flow equation.\cite{MSE} \ 
After doing so, we need no momentum integration and therefore the
computational time is reduced drastically.
We can not do Taylor expansion in momentum components $p_\mu$
because the sharp cutoff induces non-analyticity at the origin of 
the momentum space.\cite{MSE,Wilson_Kogut} \ 
Instead we expand one particle irreducible generating functional 
$\widetilde\Gamma_\Lambda$
in momentum scale $|p|=\sqrt{p_\mu p_\mu}$.
\footnote{The $M$th order approximation is
to drop all terms beyond $O(|p|^M)$ and 
the lowest order approximation coincides with the local potential
approximation of the  Wegner-Houghton equation.\cite{MSE}} 

In general we may express the cutoff function $C(q/\Lambda)$ as
$C^{-1}(q/\Lambda)=1/\theta_\varepsilon(|q|,\Lambda)-1$ in terms of 
$\theta_\varepsilon(|q|,\Lambda)$ which is a smooth regularization of
Heaviside $\theta$ function of width $\varepsilon$.
In the sharp cutoff limit $(\varepsilon\to 0)$, 
$\theta_\varepsilon(|q|,\Lambda)$ reduces to $\theta(|q|-\Lambda)$
and the evolution equation for the effective average action becomes
\be
\frac{d}{d\Lambda}\widetilde\Gamma_\Lambda[\bar\psi,\psi,\sigma,\pi]
=-\frac{1}{2}{\bf str}
\left[\frac{\delta(|q|-\Lambda)}{\gamma (q,\Lambda)}
{\hat \Gamma_\Lambda^{(2)}}\left(
1+
G{\hat \Gamma}_\Lambda^{(2)}\right)^{-1}\right],
\label{eq:Sharp_Cutoff_RGE}
\ee
where we separate the field independent full inverse propagator 
$\gamma(p,\Lambda)$ 
from the two point function
\be
\left(\widetilde\Gamma_\Lambda^{(2)}[\bar\psi,\psi,\sigma,\pi]\right)_{pp'}=
\gamma(p,\Lambda)~(2\pi)^4~\delta({ p}+{ p'})
+\left({\hat \Gamma}^{(2)}_\Lambda[\bar\psi,\psi,\sigma,\pi]\right)_{pp'},
\ee
so that ${\hat \Gamma}^{(2)}_\Lambda[0,0,0,0]=0$.\cite{MSE} \ 
In Eq.~(\ref{eq:Sharp_Cutoff_RGE}),
$G(p,\Lambda)$ is the infra-red cutoff propagator,
\be
G(p,\Lambda)=\lim_{\varepsilon\to 0}
\frac{1}{C^{-1}(p/\Lambda)+\gamma(p,\Lambda)}
=\frac{\theta(|p|-\Lambda)}{\gamma(p,\Lambda)}.
\ee
As mentioned above, in the sharp cutoff limit 
the origin of momentum space is not analytic and 
Taylor expansion in $p_\mu$ breaks down. 
Alternatively, one should expand in momentum scale.
The infra-red cutoff function is expanded in terms of absolute value of 
external momenta $|p|$ as
\be
\theta(|p+q|-\Lambda)=
\theta(q\cdot {\hat p}+|p|/2)=
\theta(q\cdot {\hat p})
+\sum_{m=1}^{\infty}\frac{1}{m!}\left(\frac{|p|}{2}\right)^m
\delta^{(m-1)}(q\cdot {\hat p}).
\ee
Here ${\hat p}_\mu$ is a unit vector parallel to $p_\mu$.
Integrating RHS of Eq.~(\ref{eq:Sharp_Cutoff_RGE}) 
with respect to the internal momenta $q$, 
one can expand it
in terms of the momentum scale $|p|$.
In finite temperature case, 
the momentum integral in Eq.~(\ref{eq:Sharp_Cutoff_RGE})
turns to the angular average over three dimensional momenta which 
are restricted to ${\bf q}^2=\Lambda^2-\omega_{(b~\mbox{{\tiny or}}~f),n}^2$ 
for each Matsubara mode.
Using the formula of angular integral,
\be
\int d^3q~\delta\left(|{\bf q}|
-\sqrt{\Lambda^2-\omega_{(b~\mbox{{\tiny or}}~f),n}^2}
\right)
f({\bf q}\cdot{\hat{\bf p}})
=2\pi(\Lambda^2-\omega_{(b~\mbox{\tiny{or}}~f),n}^2)\int_{-1}^{1}dz~f(z),
\ee
we can integrate with respect to the internal momenta ${\bf q}$ and 
expand RHS of RGE\\
~(\ref{eq:Sharp_Cutoff_RGE}) 
in terms of the external momenta $|{\bf p}|$.
The flow equation for the effective potential $V(\rho)$ 
can be derived. 
We have
\be
\frac{d}{dt} V=\beta_f+\beta_b,
\ee
where $\beta_f$ is the contribution from fermion loop graphs
and $\beta_b$ is that from boson loop graphs. 
They are expressed as;
\bea
&&\beta_f=
-\frac{T}{\pi^2}{\sum_n}'\zeta_{f,n}\ln
\left[1+\frac{2y^2\rho}
{Z_f^2\Lambda^2+(Z_{f\tiny{0}}^2-Z_f^2)~\omega_{f,n}^2-\mu^2
+2i~Z_{f\tiny{0}}~\omega_{f,n}\mu}\right],\nonumber\\
&&\beta_b=
\frac{T}{4\pi^2\CO}{\sum_n}'\zeta_{b,n}\ln \left[ Z_b
\Lambda^2+( Z_{b\tiny{0}}- Z_b)~\omega_{b,n}^2
 +V'\right]\nonumber\\
&&\qquad\quad+\frac{T}{4\pi^2\CO}{\sum_n}'\zeta_{b,n}
\ln\left[ Z_b \Lambda^2
+( Z_{b\tiny{0}}- Z_b)~\omega_{b,n}^2
 +V'+2\rho  V''\right]\nonumber\\
&&\qquad\quad-\frac{T}{4\pi^2\CO}{\sum_n}' \zeta_{b,n}
 \ln \left .\left[ Z_b \Lambda^2
+( Z_{b\tiny{0}}- Z_b)~\omega_{b,n}^2+V'
 \right]^2\right|_{\rho=0}.
\eea
where $\zeta_{(b~{\rm or}~f),n}=\sqrt{\Lambda^2\
-\omega_{(b~{\rm or}~f),n}^2}\cdot\Lambda^2$ 
and the prime $'$ operating on the effective potential $V$ denotes 
the derivative with respect to $\rho$.
$\sum'_n$ is the summation about $n$ with the condition 
$\omega_{(b~{\rm or}~f),n}^2\le\Lambda^2$.
For the wave function renormalization factor $Z_b$, we have
\footnote
{In momentum scale expansion of sharp cutoff effective average action,
there emerges a kinetic term proportional to $|{\bf p}|$.
The non-analytic dependence on momenta at ${\bf p}=0$ lead to 
a non-locality in position space. 
Since the resulting physics in the limit $\Lambda\to 0$ should not suffer
from this non-locality, it should be absorbed by a certain 
counter term. Here we simply neglect it in our approximation. } 
\bea
\frac{d}{dt} Z_b&=&
\frac{2T}{\pi^2}{\sum_n}'\frac{\zeta_{f,n}Z_f^2~y^2}
{(Z_f^2\Lambda^2+(Z_{f\tiny{0}}^2-Z_{f}^2)
~\omega_{f,n}^2-\mu^2-2iZ_{f\tiny{0}}\mu~\omega_{f,n})^2}\nonumber\\
&&\cdot\left(\frac{1}{2}-\frac{1}{3}\frac{Z_f^2(\Lambda^2-\omega_{f,n}^2)}
{Z_f^2\Lambda^2+(Z_{f\tiny{0}}^2-Z_{f}^2)
~\omega_{f,n}^2-\mu^2-2iZ_{f\tiny{0}}\mu~\omega_{f,n}}\right).
\eea
In this paper, we also approximate $ Z_{b\tiny{0}}$ as 
$ Z_{b\tiny{0}}= Z_{b}$. In the large $\CO$ limit, 
the boson fluctuations disappeared, $i.e.$, $\beta_b=0$
due to a factor $1/\CO$. 
However for finite $\CO$, 
we must incorporate the boson loop contributions, $\beta_b$. 
This is a significant difference from the large 
$\CO$ limit. In general the fermion contribution makes the effective potential 
evolve downward as cutoff is lowered, while the boson contribution
lifts the effective potential upward.
\footnote{This is understood from the signs of $\beta_f$ and $\beta_b$.}
The latter is main additional effect for finite $\CO$ 
on the RG evolution of the effective potential and hence
also on the chiral phase structure on $(T,\mu)$ plane. 
We shall ignore the corrections to fermion's $Z$ factors 
$Z_{f}$, $Z_{f\tiny{0}}$ and to yukawa coupling $y$\footnote{
Here, we do not renormalize the boson fields $\sigma,\pi$ to 
make their $Z$ factor unity, $i.e.$ $ Z_b=1$, 
since at ultra-violet cutoff, it should vanish. 
Therefore in an ordinary sense, our physical yukawa coupling runs 
due to bosonic wave-function renormalization.} as the first
step toward the exploration of the phase structure at 
finite temperature and finite chemical potential beyond the $1/\CO$ leading.
\footnote{
A similar approximation is performed in Ref.~\citen{Wett_Baryon}.}

\begin{wrapfigure}{l}{\halftext}
\epsfxsize=\halftext
\epsfbox{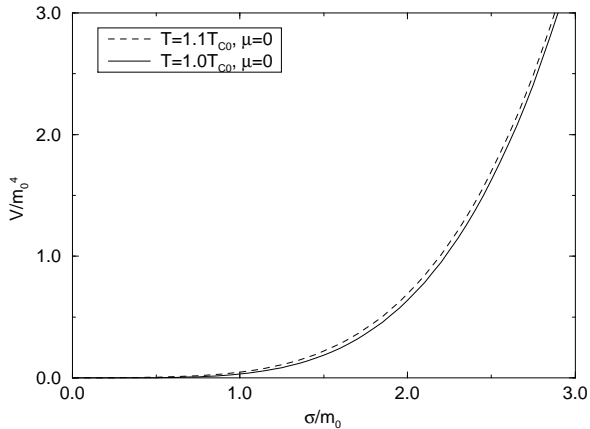}
\caption{The effective potential of four dimensional
NJL model at finite temperature.
Here, $N_{\mbox{c}}=10$. 
The effective potential and the field expectation value are 
normalized by 
the dynamical fermion mass $m_0$ at $T=\mu=0$.
\label{PotT0003N10}
}
\end{wrapfigure}
RGE for the effective potential is non-linear partial differential
equation. 
Here we do not attempt to expand it in powers of the fields 
as done in Ref.\citen{rf:comoving}, 
since if the cubic term of the field $i.e.$ $\rho^{3/2}$ 
appears in the effective potential $V(\rho)$,\cite{rf:Dolan} then  
the naive polynomial approximation of it can not work.
We discretize the RG flow equation in $t$ and $\rho$ directions 
and solve it numerically. 
We apply the extended Crank-Nicholson method\cite{Numerical_Recipes,Ogure}~
which is effective for a non-linear partial differential equation. 

By evaluating the minimum of the effective potential, we can conclude whether 
it is a broken phase or a symmetric one.
If there is a non-trivial minimum at nonzero value of $\sigma$, it is 
a broken phase. 
If the temperature (or the chemical potential) is greater than
the critical one, the effective potential has a minimum at the origin, 
{\it i.e.} the chiral symmetry is recovered.
If the temperature (or the chemical potential) is below the critical one,
the effective potential has a nontrivial minimum.
More precisely, in the broken phase, the effective potential evolves
toward the so-called `convex' one.\cite{CEP,convex}
Some results of the effective potentials are shown in Fig.~\ref{PotT0003N10}.
\begin{figure}
\epsfxsize=0.65\textwidth
\centerline{\epsfbox{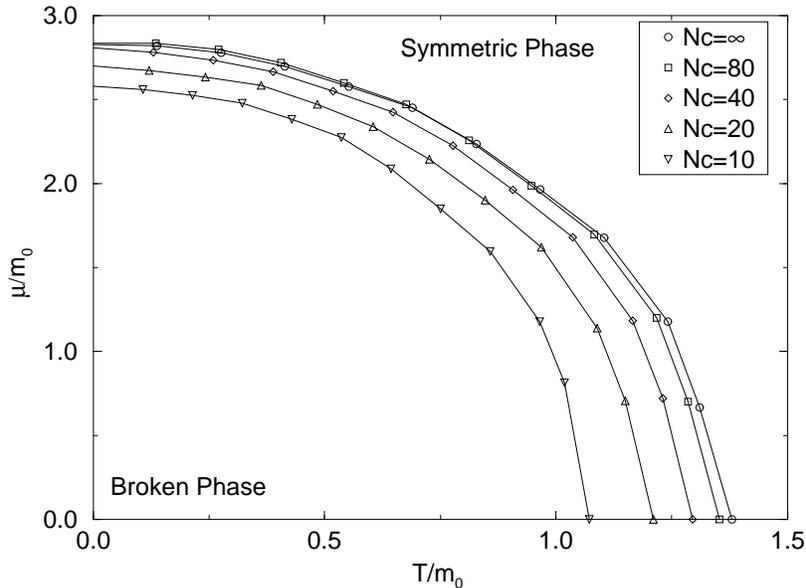}}
\caption{$N_{\mbox{c}}$ dependence of
the chiral phase structure 
of the four dimensional NJL model
at finite temperature and chemical potential.
In each $N_{\mbox{c}}$, the phase transition is 
of second order or very weak first order all along the critical line.
The temperature and the chemical potential are normalized by 
the dynamical fermion mass $m_0$ at $T=\mu=0$.
\label{TDplaneD4}
}
\end{figure}
\begin{table}
\caption{The critical temperature $T_{\mbox{c0}}$
and the chemical potential $\mu_{\mbox{c0}}$ 
of the four dimensional NJL model.}
\label{table:1}
\begin{center}
\begin{tabular}{c|cccccccc}
\hline 
\hline 
$\CO$ & 
$10$ & $20$ & $40$ & $80$ & $\infty$ \\
\hline
$T_{\mbox{c\tiny{0}}}/m_0$ & 
$1.07$ & $1.21$ & $1.30$ & $1.35$ & $1.38$ \\
$\mu_{\mbox{c\tiny{0}}}/m_0$ &
$2.58$ & $2.70$ & $2.81$ & $2.83$ & $2.83$ \\
$\mu_{\mbox{c\tiny{0}}}/T_{\mbox{c\tiny{0}}}$ &
$2.41$ & $2.23$ & $2.16$ & $2.10$ & $2.05$ \\
\hline
\end{tabular}
\end{center}
\end{table}

Taking into account of the discussions in the previous section,
we take an initial condition of RG flow equations 
at $\Lambda_0=333~T_{\mbox{c\tiny{0}}}$.
Indeed, with this condition, the critical line almost coincided with 
that from SDE in three dimensional GN model 
for $\CO=\infty$ (See Fig.~\ref{TDplaneD3I}).
The chiral phase diagram on $(T,\mu)$ plane is shown 
in Fig.~\ref{TDplaneD4}. 
Here, the temperature and the chemical potential 
are normalized by the dynamical fermion mass $m_0=y<\sigma>$ at 
zero temperature and zero chemical potential. 
In each $\CO$ case, the phase transition is found to be of second order
or very weak first order all along the critical line
from th shape of the effective potentials at the critical points.
For example, an effective potential at a critical point 
for $\CO=10$ is shown in Fig.~\ref{PotT0003N10}.
$\CO$ dependence of the critical line on $(T,\mu)$ plane is not small.
The critical temperature and the critical chemical potential 
become small in
unit of $m_0$ if $\CO$ is lowered.
This is due to the boson fluctuations existing for finite $\CO$. 
Intuitively, we can understand it as follows.
The fermionic negative corrections $(\beta_f)$ to the effective potential 
are suppressed by
a high temperature and/or a large chemical potential, 
but the bosonic positive ones $(\beta_b)$ are less suppressed 
due to the existence of Matsubara zero mode. 
Hence if the bosonic part
becomes large, that is, the case of small $\CO$, the chiral symmetry 
is restored at lower temperature (smaller chemical potential) 
in comparison with
the large $\CO$ case.
$\CO$ dependence of the critical temperature at $\mu=0$ 
($T_{\mbox{c\tiny{0}}}$) as well as 
that of the critical chemical potential at $T=0$ 
($\mu_{\mbox{c\tiny{0}}}$) are shown in 
Table \ref{table:1}. 
\footnote{The numerical calculations for $\CO <10$ need much finer
mesh in $t$ direction because $\beta_b$ becomes large.
This will be a subject in the forthcoming paper.}

\section{Summary and discussion}

We performed a non-perturbative analysis by the (Wilsonian) non-perturbative 
renormalization group in NJL/GN model at finite temperature and 
chemical potential and explored chiral phase structure on $(T,\mu)$
plane.

First we explored the phase structure of the GN
model at finite 
temperature and chemical potential in the large $\CO$ limit. 
We showed that consistent results with those of SDE 
could be obtained in the framework of NPRG method.
We investigated how large ultra-violet cutoff $\Lambda_0$ 
is needed to obtain scaled quantities.
We could calculate scaled critical temperature/chemical potential
by setting the initial condition at $\Lambda_0=333~T_{\mbox{c\tiny{0}}}$
and they had little cutoff scheme dependence.

Second we improved the approximation beyond the $1/\CO$ leading
in four dimensional NJL model.
For finite $\CO$, the bosonic fluctuations have to be incorporated. 
In \S\,4, we introduced the mesonic auxiliary fields $\sigma,\pi$, 
and derived the flow equation for the effective potential of 
these fields $V(\sigma^2+\pi^2)$. RG flow equation is then a non-linear 
partial differential equation for this potential. 
Taking account of 
the possibility of the first order phase transition like 
Fig.\ref{Pot_1st} and the cubic term $\rho^{3/2}$,\cite{rf:Dolan} 
we solved the partial differential equation for 
$V(\sigma^2+\pi^2)$ without a polynomial expansion employed in 
Ref.\citen{rf:comoving}. 
For several $\CO$, we obtained the chiral phase structure 
on $(T,\mu)$ plane. 
For $\CO =10\sim\infty$, the phase transition is second order
or very weak first order
all along the critical line.
The critical temperatures/chemical potentials 
depend on $\CO$ largely, and become small
as $\CO$ is decreased.
For $\CO=10$, $T_{\mbox{c\tiny{0}}}$ is $78 \%$ of that for $\CO=\infty$. 

Generalization of this analysis to more realistic models, QCD 
seems to be straightforward, except for the treatment of the gauge invariance. 
By introducing the instanton induced multi-fermi operator,\cite{instanton} 
NPRG method can be also applied to the color-superconductor.\cite{rf:CSC}

\section*{Acknowledgments}
We would like to thank W.Souma for informative discussions 
on numerical computation,
and K-I.Aoki, T.Suzuki, and H.Terao for useful comments.
Part of numerical computations in this work were carried out in the 
Yukawa Institute Computer Facility.
\appendix
\section{Threshold functions} 
In this appendix we show the explicit form of the threshold functions 
$I(a,b~;{\widehat T},{\widehat\mu})$ appeared in Eq.(\ref{eq:rgs3}).
In $d$ dimensions, the threshold function at zero temperature and zero chemical
potential is
\be
I(a,b~;0,0)
=\frac{1}{2^{d-1}\pi^{d/2}\Gamma(d/2)}\int^{\infty}_{0}
dq~q^{d-3} 2^{(d/2+2)} a~b~q^{2b} (1-f^2(q))f^2(q).
\label{eq:A1}
\ee
For finite temperature and/or chemical potential, the threshold function
can be obtained by the replacement 
$\int dq^d/(2\pi)^d \to {\widehat T}\sum_n \int dq^{d-1}/(2\pi)^{d-1}$ and 
$q_0 \to \widehat\omega_{f,n}-i\widehat\mu
=(2n+1)\pi {\widehat T}-i\widehat\mu$,
\bea
&&I(a,b~;{\widehat T},{\widehat\mu})
=\frac{1}{2^{d-2}\pi^{(d-1)/2}\Gamma((d-1)/2)}
{\widehat T}\sum_{n=0}^{\infty}\int^{\infty}_{0}
dq~q^{d-2} 2^{(d-1)/2+3} a~b~{(\widehat\omega_{f,n}^2+q^2)}^b\nonumber\\ 
&&\qquad\qquad\quad\cdot(1-f^2(q))f^2(q)
\frac{\widehat\omega_{f,n}^2-\widehat\mu^2+q^2}
{\left\{\widehat\omega_{f,n}^2+(q+{\widehat\mu})^2\right\}
 \left\{\widehat\omega_{f,n}^2+(q-{\widehat\mu})^2\right\}}.
\label{eq:A2}
\eea
${\widehat T}\to 0$ and $\widehat\mu\to 0$ limit 
of Eq.(\ref{eq:A2}) coincides with 
Eq.(\ref{eq:A1}).

\section{Renormalization and the continuum limit at $T,\mu\ne 0$}
Let us discuss renormalization and the continuum limit 
at finite temperature and finite chemical potential
in this appendix.
In the large $\CO$ limit, the RG $\beta$ functions are given by,
\bea
&&\dt{\widehat\GS}=-(d-2)~{\widehat\GS}+I(\T,\M)~{\widehat\GS}^2,\\
&&\dt\T=\T,\\
&&\dt\M=\M.
\eea
Fixed points of the above RG equations at $\T=\M=0$ are 
the Gaussian fixed point 
${\widehat\GS}=0$ and Gross-Neveu point 
${\widehat\GS}=(d-2)/I(\T=0,\M=0)\equiv{\widehat\GS}^\star$. 
Note that, $\T=\infty$ and/or 
$\M=\infty$ are also fixed points. If $\T=\infty$ and/or $\M=\infty$ then
only Gaussian fixed point can be found for the finite ${\widehat\GS}$, 
since $I(\T,\M)$ vanishes. 
The difference from an ordinary zero temperature 
and zero chemical potential
field theory is $\T,\M$ dependence of the threshold function $I$. 
As will be explained below, it does not affect the `renormalization'.

We linearize the RG equations around the GN point.
\footnote{For the infinitesimal neighborhood of the fixed point, 
this linearization is valid.}\  
Letting ${\widehat\GS}={\widehat\GS}^\star+\delta{\widehat \GS}$, we have,
\bea
&&\dt\delta{\widehat \GS}
=-(d-2)~\delta{\widehat \GS}+2I_0~{\widehat\GS}^\star~\delta{\widehat \GS}
+\left(\T I^T_0+\M I^\mu_0\right){\widehat\GS}^{\star 2}\\
&&\dt\T=\T,\\
&&\dt\M=\M,
\eea
where $I_0=I(\T=0,\M=0)$, $I^T_0=(\D I/\D\T)(\T=0,\M=0)$ and 
$I^\mu_0=(\D I/\D\M)(\T=0,\M=0)$. By Lorentz symmetry at $\T=\M=0$, 
$I^\mu_0=0$. We also find $I^T_0=0$, 
since the difference between $I(\T,0)$ and $I_0$ is 
${\cal O}(\T^2)$.
\footnote{
$I(\T,0)$ is like an approximation of the integral $I_0$ by
a histogram, band width of which is $2\pi {\widehat T}$.
As a trapezoidal rule estimate differs from an integral
by an order of the square of the width,
\cite{Numerical_Recipes}
$I(\T,0)$ also differs from $I_0$ by ${\cal O}(\T^2)$.}
Consequently, we find,
\be
\dt
\left(\begin{array}{c}
\delta{\widehat \GS}\\
\T\\
\M
\end{array}\right)
=\left(\begin{array}{ccc}
-(d-2)+2I_0{\widehat\GS}^\star  &   & {\bf 0}\\
        & 1 &        \\
{\bf 0} &   &  1
\end{array}
\right)
\left(\begin{array}{c}
\delta{\widehat \GS}\\
\T\\
\M
\end{array}\right).
\ee
The eigenvalues are found to be $\nu_1=d-2,\nu_2=\nu_3=1$. 
Here we used ${\widehat\GS}^\star=(d-2)/I_0$. 
We derive the relations between the bare coupling and the physical
quantities using an ordinary procedure.
Let us focus on the infinitesimal neighborhood of GN point, 
$i.e.$ we consider 
the continuum limit of the theory $\Lambda_0\to\infty$. 
The critical temperature $T_{\rm c}$ 
is completely determined by 
mutually independent three variables: the bare four-fermi coupling 
$\delta{\widehat \GS}$, bare chemical potential $\M_0$ and 
the ultra-violet cutoff $\Lambda_0$. 
The extra $\M_0$ dependence of the critical temperature $i.e.$
$T_{\rm c}(\delta{\widehat \GS},\Lambda_0;\M_0)$ does not affect to 
the following considerations. 
Now it is convenient to think that the critical temperature 
depends on the `bare' chemical potential $\M_0$ through the fixed 
as well as RG invariant dimensionless ratio $\M_0/\T_0$. 
As everyone knows, the RG preserves the physical quantities. 
Therefore, we can realize by solving 
the RG equations from $\Lambda_0$ to $\Lambda=\lambda\Lambda_0$ that 
the critical temperature satisfies,
\be
T_{\rm c}(\delta{\widehat \GS},\Lambda_0;\M_0/\T_0)=
T_{\rm c}(\lambda^{-(d-2)}\delta{\widehat \GS},\lambda\Lambda_0;\M_0/\T_0),
\label{eq:scale}
\ee
where we used the solution,
\be
\delta{\widehat \GS}=C {\rm e}^{(d-2)t}=C(\Lambda_0/\Lambda)^{d-2}.
\ee
Here $C$ is the `bare' $\delta{\widehat \GS}(\Lambda_0)$. 
The relation (\ref{eq:scale}) can be rewritten by the dimensionless 
critical temperature $\T_{\rm c}(\delta{\widehat \GS};\M_0/\T_0)
=T_{\rm c}(\delta{\widehat \GS},\Lambda_0;\M_0/\T_0)/\Lambda_0$
\footnote{%
By dimensional analysis, $\T_{\rm c}$ is independent of $\Lambda_0$.}
as
\be
\T_{\rm c}(\delta{\widehat \GS};\M_0/\T_0)=
\lambda\T_{\rm c}(\lambda^{-(d-2)}\delta{\widehat \GS};\M_0/\T_0).
\ee
Hence,
\be
\T_{\rm c}(\delta{\widehat \GS};\M_0/\T_0)\propto
\delta{\widehat \GS}^{1/(d-2)}.
\ee
If one starts from the temperature dependent critical chemical potential, 
then one finds,
\be
\M_{\rm c}(\delta{\widehat \GS};\M_0/\T_0)\propto
\delta{\widehat \GS}^{1/(d-2)}.
\ee

One can also find a similar relation of the fermion dynamical mass 
at $T=\mu=0$, $m_0\propto\delta{\widehat \GS}^{1/(d-2)}$. 
We can `renormalize' 
these quantities by taking 
$\delta{\widehat \GS}(\Lambda_0)=(M/\Lambda_0)^{(d-2)}$,
where $M$ is some finite reference mass scale. Letting $\Lambda_0\to
\infty$, we find the continuum limit of GN model at finite temperature 
and/or finite chemical potential. 

The above observations can be straightforwardly generalized to other models.

\end{document}